\documentclass[aps,pra,twocolumn,showpacs,floatfix,superscriptaddress]{revtex4}

\usepackage{amsmath,float,graphicx,subfigure,epstopdf,color}

\begin{document}
\title{Density wave propagation in a two-dimensional random dimer potential:
 from a single to a bipartite square lattice}
\author{P. Capuzzi}
\email{capuzzi@df.uba.ar}
\affiliation{Universidad de
  Buenos Aires, Facultad de Ciencias Exactas y Naturales, Departamento
  de F\'{i}sica, Buenos Aires, Argentina}
\affiliation{IFIBA,
  CONICET-UBA, Pabell\'on 1, Ciudad Universitaria, 1428 Buenos Aires,
  Argentina }
\author{P. Vignolo}
\email{Patrizia.Vignolo@inphyni.cnrs.fr}
\affiliation{Universit\'e C\^ote d'Azur, CNRS, Institut de Physique de Nice,
1361 route des Lucioles
06560 Valbonne, France}
\begin{abstract}
  We study the propagation of a density perturbation in a weakly
  interacting boson gas confined on a lattice and in the presence of
  square dimerized impurities. Such a two-dimensional random-dimer
  model (2D-DRDM), previously introduced in [Capuzzi {\it et al.},
  Phys. Rev. A {\bf 92}, 053622 (2015)], is the disorder transition
  from a single square lattice, where impurities are absent, to a
  bipartite square lattice, where the number of impurities is maximum
  and coincides with half the number of lattice sites.  We show that
  disorder correlations can play a crucial role in the dynamics for a
  broad range of parameters by allowing density fluctuations to
  propagate in the 2D-DRDM lattice, even in the limit of strong
  disorder.  In such a regime, the propagation speed depends on the
  percentage of impurities, interpolating between the speed in a
  single monoperiodic lattice and that in a bipartite one.
\end{abstract}
\pacs{}
\maketitle

\section{\label{sec:intro}Introduction}
Disordered two-dimensional (2D) systems in the absence of interactions
and disorder correlations are insulating, as demonstrated in the
seminal work on the scaling theory of localization by Abrahams,
Anderson, Licciardello and Ramakrishnan \cite{Ande79}.  The concept of
correlations for a random potential $V(r)$ is related to the behaviour
of the correlation function $\overline{V(r)V(r')}=f(r-r')$ averaged
over all the disorder configurations, where the absence of
correlations corresponds to $f(r-r')=\delta(r-r')$.  However, like
disorder, potential correlations and interactions are almost
unavoidable in physics, breaking the validity of the scaling theory
\cite{Ande79,and80,tho82}.  In particular, it is well established that
short-range correlations, i.e. those with $f(r-r')\rightarrow 0$ for
$|r-r'|$ greater than few lattice spacings, can induce delocalized
states \cite{Hilke1994,Hilke2003,Capuzzi2015,Naether2015} or states
that are extended over large distances \cite{Kuhn07,Miniatura2009};
while long-range correlations may cause the absence of localization
\cite{mou04,dos07,mou08,mou10b}.  Interactions can induce a
glass-superfluid transition \cite{Fisher1989} or induced many-body
localization at finite temperature
\cite{Basko2006,Basko2007,Fleury2008}.  Moreover, for weakly
interacting systems, correlated disorder can shift the onset of
superfluidity \cite{Pilati2010,Plisson2011,Allard2012,Carleo2013}, or
enhance superfluidity itself, even in the presence of strong disorder
\cite{Capuzzi2015}.  This has been shown for the two-dimensional Dual
Random Dimer Model (2D-DRDM) \cite{Capuzzi2015}, that, analogously to
the well-known one-dimensional (1D) model
\cite{Phil90,schaff10,Larcher2013}, is a tight-binding model
characterized by correlated impurities that become ``transparent''at a
given resonance energy, like identical Fabry-Perot cavities.  If the
Hamiltonian parameters are tuned so that the resonance energy matches
the ground-state energy, the ground state is not affected by the
disorder, even in the presence of weak interactions. The density
homogenization induced by the resonance drives the superfluid fraction
\cite{Capuzzi2015}. This happens at the ground-state energy, but, as
soon as the system is perturbed, higher energy states are involved,
and it is not straightforward to derive the response of the system.
Indeed, strictly-speaking, the resonant energy in the absence of
interactions is only one (or few \cite{Larcher2013}), so that all the
other states with different energy should be localized, even if a part
of them ($\mathcal{N}^{1/2}$ in 1D, $\mathcal{N}$ being the number of
states \cite{Phil90}) is expected to be localized on the entire system
length.

In this work we study the transport of an initial ring-shaped density
perturbation in a weakly interacting boson gas confined on a 2D-DRDM
lattice, and we compare it with the case of a fully uncorrelated
random lattice (UN-RAND).  We analyze both the shape of the
perturbation and its propagation speed as a function of the disorder
properties. Far from the resonance condition, the density fluctuation
distorts as it travels through the lattice irrespective of the model
disorder. However, close to the resonance condition, in the 2D-DRDM,
the density perturbation travels through the system without broadening
and with a well-defined speed.  The propagation speed depends on the
percentage of impurities and its value is between the speed of a
density perturbation in a single square monoperiodic (MP) lattice and
that in a bipartite (BP) one composed by two interlacing square
sublattices.  This shows that the main role of ``resonant''random
dimer impurities in a MP lattice or vacancies in a PB one is to drive
the value of the propagation speed during the transport of a density
excitation.

The paper is organized as follows. In Sec. \ref{sec-model} we review
the features of the 2D-DRDM, including the location of the
single-particle energy resonance and its effect on the spectral
function for non-interacting particles. The results for the density
wave propagation in a weakly interacting many-particle system are
presented in Sec. \ref{sec:results}. In this section, we compare the
numerical results obtained {\it via} the dynamical equations using a
Gutzwiller approach, with the speed for a density perturbation in a MP
lattice and a BP one, obtained within a Bogoliubov approach strictly
valid for a pure Bose-Einstein condensate. We show that when the
resonance condition is fulfilled, there is a well-defined speed for
the density propagation and that the density wave packet is not
broadened by the disorder during its propagation.  Concluding remarks
and perspectives are given in Sec. \ref{sec:concl}.

\section{The system}
\label{sec-model}
\begin{figure*}
\begin{center}
\includegraphics[width=0.9\linewidth]{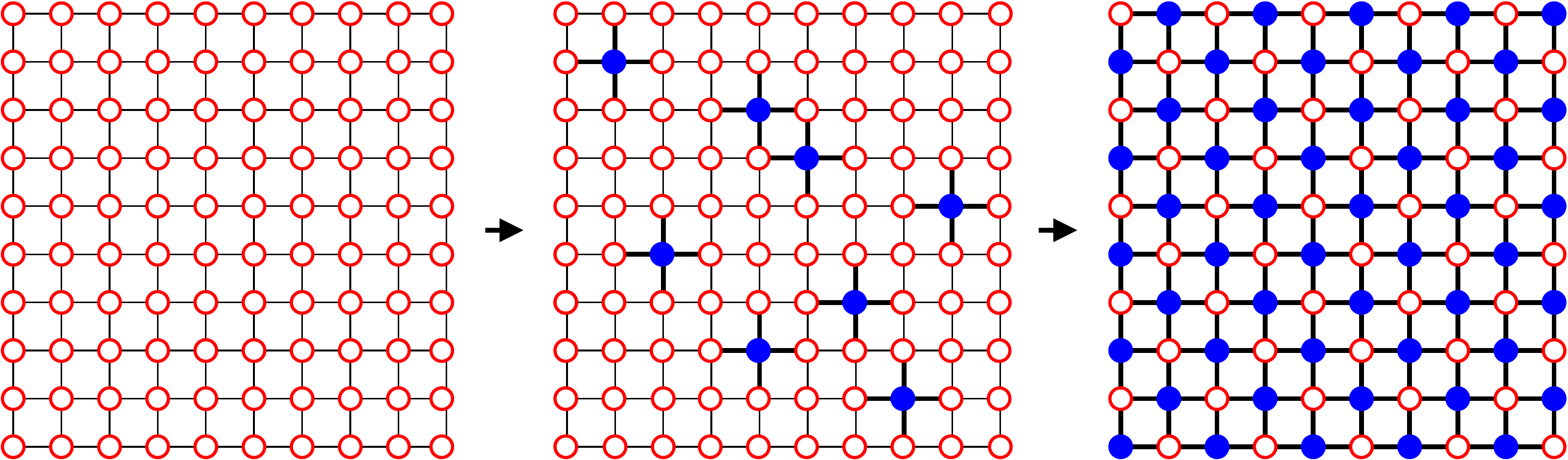}
\caption{\label{fig1} Schematic representation of the 2D-DRDM
  (middle panel) as the disordered interpolation between the
  monoperiodic 2D lattice (left panel) and the bipartite one
  (right panel).}
\end{center}
\end{figure*}

The 2D-DRDM is a single-particle tight-binding model, characterized by
``isolated'' on-site impurities that locally modify the hopping
probability (middle panel of Fig. \ref{fig1}).  The sites are arranged
in a two-dimensional square lattice of size $L\times L$ and spacing
$a$. The system is described by the Hamiltonian in the site basis
$\{|i\rangle\}$
\begin{equation}
  H=-\sum_{\langle ij\rangle}t_{ij}(|i\rangle\langle j|+|j\rangle\langle i|)
  +\sum_{i=1}^N\varepsilon_i
  |i\rangle\langle i|,
\end{equation}
where $N=L^2$ is the number of sites and $\langle ij\rangle$ denotes
the sum over first-neighbor sites.  Here, $\varepsilon_i$ are the
on-site energies that can be 0 or $\Delta$ in the absence or presence
of an impurity, respectively, and $t_{ij}$ are the first-neighbor
hopping terms that can take two values: $t$ between two empty sites
and $t'$ between an empty site and a site hosting an impurity.
The fact that the impurities cannot be next-neighbours introduces
short-range correlations in the disorder.
Such a
potential could be realized by dipolar impurities pinned at the minima
of a lattice potential \cite{Larcher2013}.  If the percentage $p$ of
impurities is zero, the lattice is MP with site energies equal to 0
and all hopping parameters equal to $t$ (left panel of
Fig. \ref{fig1}), while if $p=0.5$, the lattice is BP with the site
energies 0 and $\Delta$ distributed in a checkerboard configuration
and all hopping parameters equal to $t'$ (right panel of
Fig. \ref{fig1}). More impurities cannot be accommodated in the 2D-DRDM
lattice so that $p=0.5$ is the maximum value that can be
attained. Furthermore, since impurities and vacancies have the same
role, the most disordered configuration corresponds to the middle
region, at $p=0.25$.
Therefore, by varying $p$ the 2D-DRDM lattice can be seen as a
disorder-mediated crossover between a MP and a BP lattice.

\begin{figure*}
  \begin{center}    
  \includegraphics[width=0.4\linewidth]{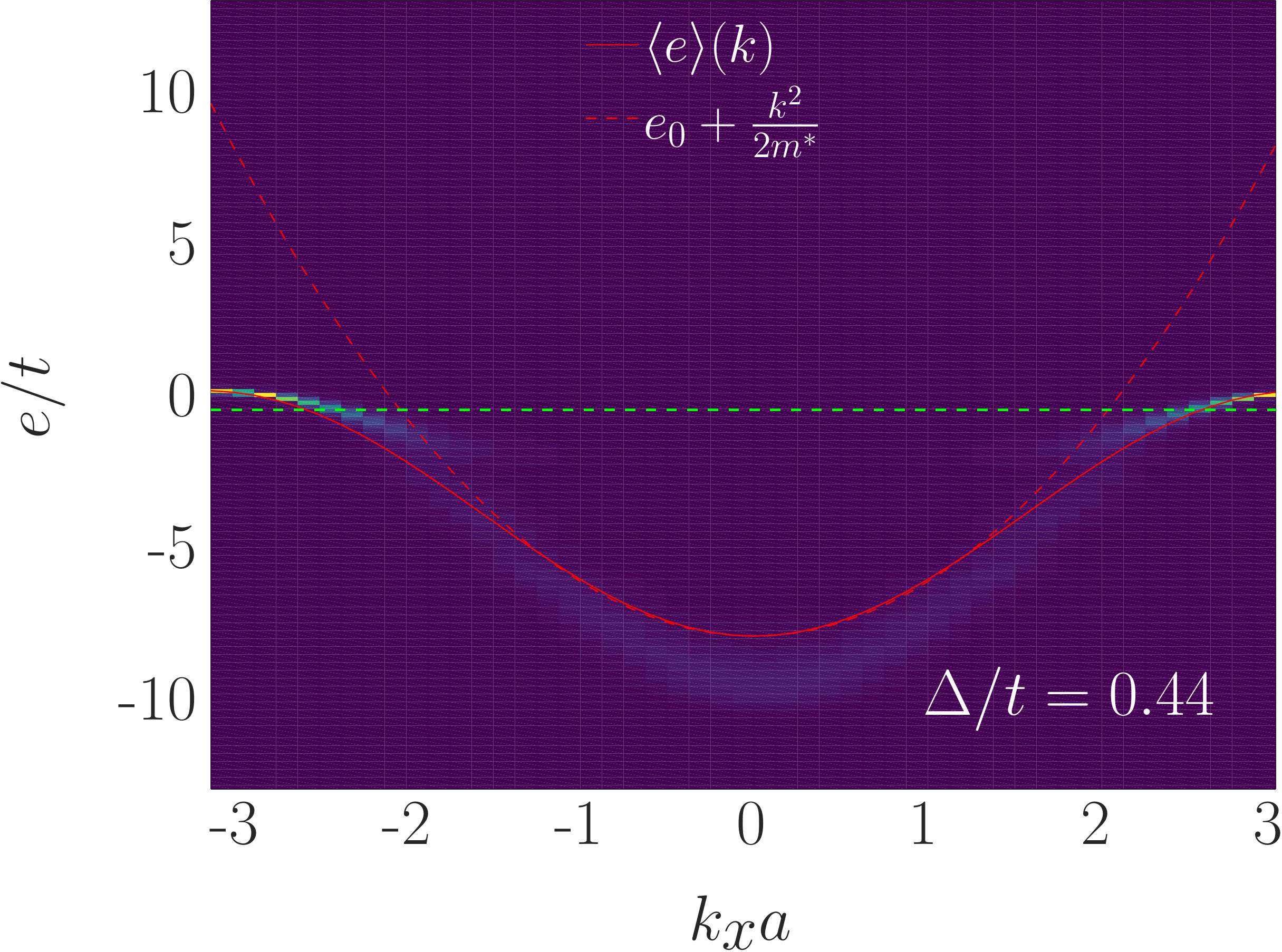}
  \includegraphics[width=0.4\linewidth]{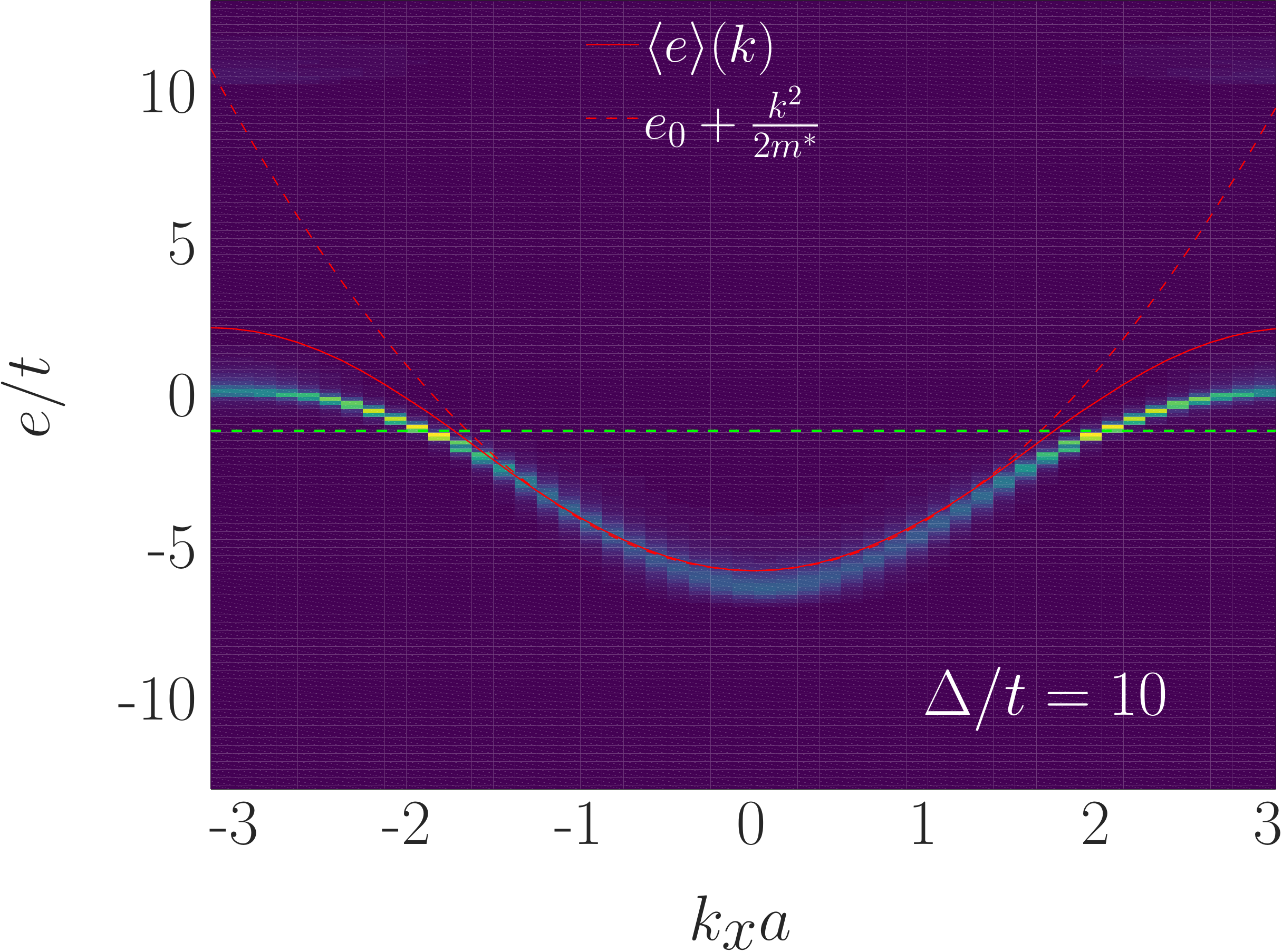}\\
  \includegraphics[width=0.4\linewidth]{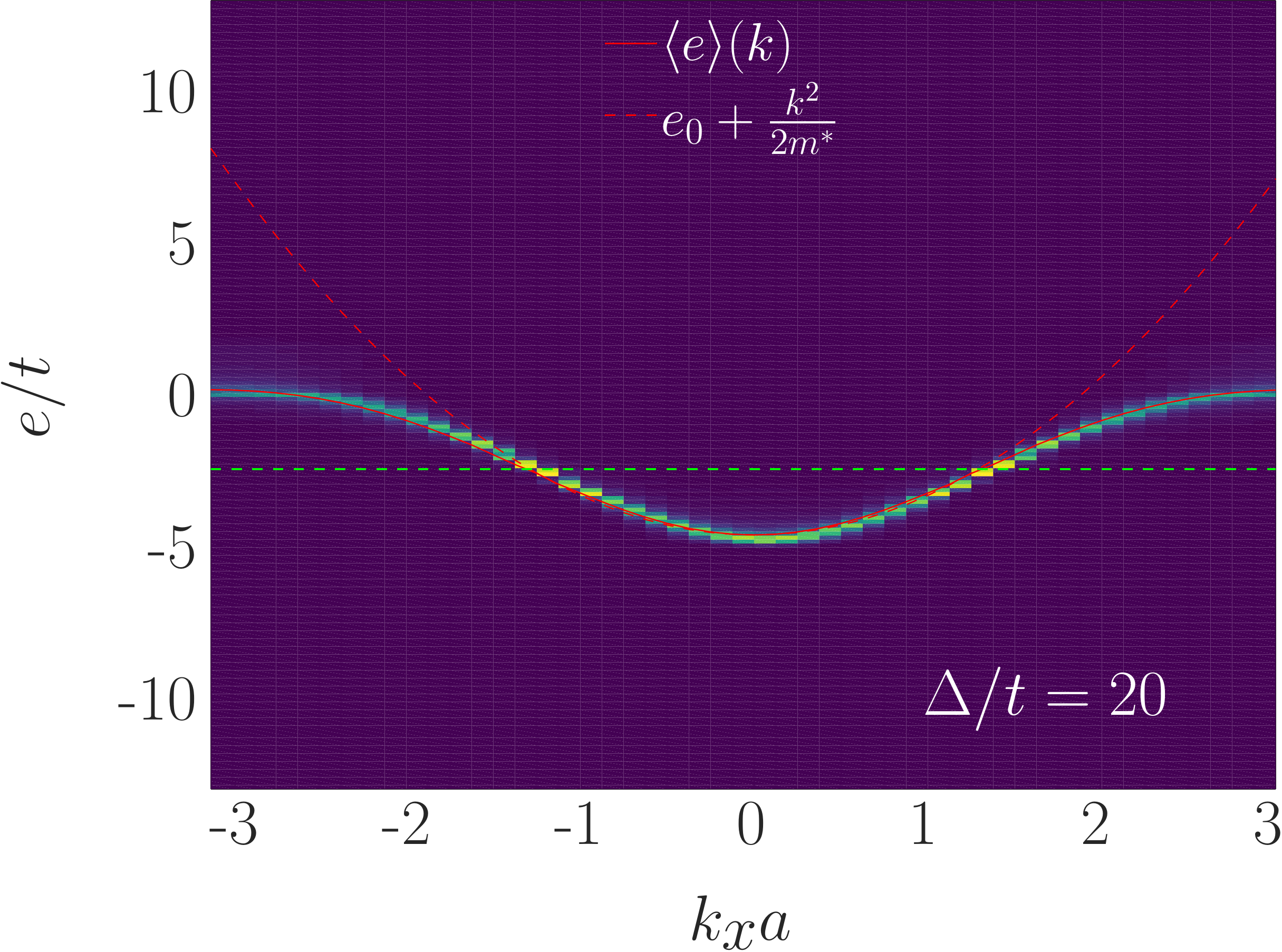}
  \includegraphics[width=0.4\linewidth]{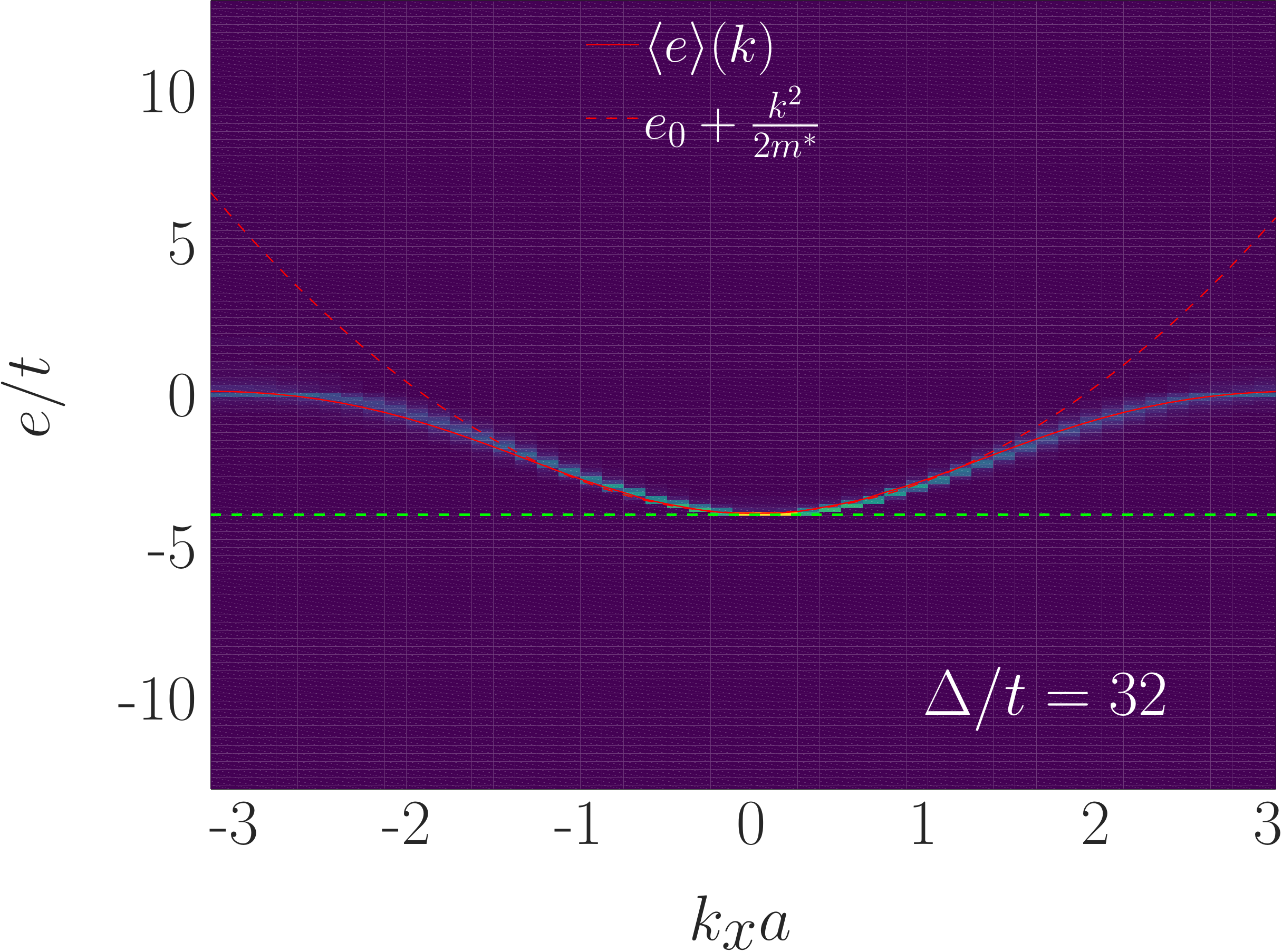}
  \caption{\label{fig-spettro} Disorder-averaged spectral function
    $\mathcal{A}(\boldsymbol{k},e)$ for a 2D-DRDM square lattice with
    $t'=3t$, $p=0.25$ and $k_y=0$ as a function of $k_x$ for several
    values of $\Delta$.  The horizontal dashed lines mark the
    resonance energies $E_\text{res}$. Continuous red lines correspond
    to $\langle e\rangle(\boldsymbol{k})$, [Eq. (\ref{disprel})] and
    the dashed red lines correspond to a quadratic fitting.  The
    different panels correspond to $\Delta/t=0.44$, 10, 20 and 32.}
\end{center}
\end{figure*}

With the aim to understand the role of the impurity structure, one can
focus on the case of a single impurity in the lattice.  Let's call $B$
the subspace defined by the impurity and $A$ the remaining lattice
subspace, composed by the remaining $N-1$ lattice sites.
We have previously shown in Ref.\ \cite{Capuzzi2015} that
it exists an energy $E_{\text{res}}$ where the Green's function in the
$A$ subspace, $G_A(E)=\langle A|(E-H)^{-1}|A\rangle$, is
the same in the presence and in the absence
of an impurity. This implies that the impurity becomes ``transparent''
at this resonance energy $E_{\text{res}}$ fulfilling
\begin{equation}
\dfrac{t^2}{E_{\text{res}}}=\dfrac{(t')^2}{E_{\text{res}}-\Delta}.
\label{eres}
\end{equation}
Therefore, at the energy $E_{\text{res}}=-t^2\Delta/((t')^2-t^2)$, the
system is not perturbed by the presence of the impurities so that
states remain delocalized over the whole system. In the case of the 1D
DRDM, it was shown in \cite{Phil90} that the number of states that are
unperturbed and extended over the entire system scales as
$\sqrt{N}$. Such a behavior can also be expected to occur for states
around $E_{\text{res}}$ in the 2D-DRDM as confirmed in
Fig. \ref{fig-spettro} where we plot the disorder-averaged spectral
function $\mathcal{A}(\boldsymbol{k},e)$ defined by
\begin{equation}
  \mathcal{A}(\boldsymbol{k},e)=\overline{\langle\boldsymbol{k}|\delta(H-e)|\boldsymbol{k}\rangle},
\end{equation} 
where $\overline{\cdots}$ denotes the average over different disorder
realizations and $|\boldsymbol{k}\rangle$ is a momentum eigenstate.
Hereafter, in our numerical calculations we shall consider a square
lattice with side $L=50$, totaling $N=2500$ sites, and open boundary
conditions. The results depicted in Fig.\ \ref{fig-spettro} correspond
to 100--500 realizations of the disorder with a percentage $p=0.25$.
The spectral function $\mathcal{A}(\boldsymbol{k},e)$ is essentially
nonzero along the average dispersion relation
\begin{equation}
\langle e\rangle(\boldsymbol{k})=\dfrac{\int \mathcal{A}(\boldsymbol{k},e)
\,e\,{\rm d}e}{\int \mathcal{A}(\boldsymbol{k},e)\,{\rm d}e},
\label{disprel}
\end{equation}
but its spread in energy strongly depends on the resonance
condition. As shown in Fig. \ref{fig-spettro}, the spectral density is
well represented by a single energy peak at $\langle e\rangle(k)$ only
around $E_{\text{res}}$ : $E_{\text{res}}=-0.55 t$ for $\Delta=0.44 t$
(first panel of Fig. \ref{fig-spettro}), $E_{\text{res}}=-1.25t$ for
$\Delta=10t$ (second panel), $E_{\text{res}}=-2.5t$ for $\Delta=20t$
(third panel), and $E_{\text{res}}=-4t$ for
$\Delta=32t$ (fourth panel).
By choosing the resonance energy at the ground-state energy
$E_{\text{GS}}^{\text{MP}}$, namely, at the bottom of the energy band
($k=0$), the energy excitations become largely unaffected by the
disorder and therefore one could expect that the long wavelength
density perturbations, i.e. the sound waves, in a weakly interacting
system will be well defined.  The condition
$E_{\text{res}}=E_{\text{GS}}^{\rm MP}$, $E_{\text{GS}}^{\rm MP}=-4t$
being the MP ground-state energy, sets the value of $\Delta$ as a
function of $t'$ and $t$:
\begin{equation}
  \Delta_{\text{res}}=4t[(t'/t)^2-1].
  \label{cond}
\end{equation}
It is worthwhile noticing that if $\Delta=\Delta_{\text{res}}$, the MP
ground-state energy, $E_{\text{GS}}^{\rm MP}=-4t$, and the BP one,
$E_{\text{GS}}^{\rm BP}=\frac{1}{2}[\Delta-\sqrt{\Delta^2+64 (t')^2}]$
coincide.  Indeed, one could start from the BP lattice, and introduce
the disorder as vacancies, the 2D-DRDM being a disordered BP lattice.
By calculating the Green's function in the presence and in the absence
of a single vacancy, and by imposing that $E=E_{\text{GS}}^{\rm BP}$,
one obtain exactly the same resonance condition (\ref{cond}).  In the
presence of weak interactions, we have previously shown that the
resonance is shifted to lower values of $\Delta$ \cite{Capuzzi2015}
and it is accompanied by a minimization of the density fluctuations
and enhancement of the superfluid fraction.

\section{\label{sec:results}Density wave propagation}
In order to describe the propagation of a density fluctuation we
include interactions into the system. An interacting boson gas
confined in an optical lattice can be described by the Bose-Hubbard
Hamiltonian, which in the grand canonical ensemble reads
\begin{multline}
  H_{\text{BH}}=-\sum_{ij}t_{ij}(\hat a_i^\dagger\hat a_j+\hat a_j^\dagger\hat a_i)
  -\sum_i(\mu-\varepsilon_i)\hat n_i\\+\dfrac{U}{2}\sum_i\hat n_i(\hat n_i-1),
  \label{BH}
\end{multline}
where $\hat a_i^\dagger$ is the creation operator defined at the
lattice site $i$ and $\hat n_i=\hat a_i^\dagger\hat a_i$. As in the
case of the single-particle model, the hopping parameters $t_{ij}$ are
chosen to describe either the 2D-DRDM or the UN-RAND lattice. The
parameter $U$ is the interparticle on-site interaction strength, and
$\mu$ denotes the chemical potential fixing the average number of
bosons. We study the dynamics governed by the Hamiltonian (\ref{BH})
using the time-dependent Gutzwiller ansatz for the wave function
\cite{Rokhsar1991,Krauth1992a,Jaksch1998}
\begin{equation}
|\Phi(\tau)\rangle=\prod_i^{L\times L}\sum_{n_i}f_i(n_i,\tau)|n_i\rangle,
\end{equation}
where $f_i(n,\tau)$ is the probability amplitude of finding $n$
particles on site $i$ at time $\tau$. The Gutzwiller approach has been
previously used to describe the superfluid-insulator transition and
the stability of bosons in an optical lattice with and without random
local impurities \cite{Altman2005,Buonsante2009,Powell2011}. It allows
interpolating between the deep superfluid and the Mott insulating
regimes.  The dynamical equations obeyed by the amplitudes
$f_i(n,\tau)$ can be obtained variationally by extremizing the action
$\mathcal{S}\left\{f_i(n,\tau),f_i^*(n,\tau)\right\}=\int_{\tau_1}^{\tau_2}d\tau
\mathcal{L}$, with
\begin{equation}
  \mathcal{L}=\frac{i\hbar}{2}\left( \langle \Phi(\tau)|\dot\Phi(\tau)\rangle
  - \langle \dot\Phi(\tau)|\Phi(\tau)\rangle\right) - \langle \dot\Phi(\tau)|H_{\text{BH}}|\dot\Phi(\tau)\rangle.
\end{equation}
Such a procedure allows us to study the time evolution of the system
at an affordable computational cost. However, being a grand canonical
description, there is no guarantee that the
expectation value of the number operator will remain constant in
time. In our calculations, we have verified that particle conservation
occurs typically with a relative accuracy of $10^{-4}$, so that more
demanding number conserving approaches
\cite{Schachenmayer2011,Peotta2013,Shimizu2018a} are not needed.

To probe the effect of the disorder correlations on the transport
properties we excite a density wave at the center of the lattice and
study its propagation. Such a density wave is constructed by solving
the stationary problem of the disordered system subject to an
additional Gaussian potential that shifts the on-site energies
$\varepsilon_i$ by
\begin{equation}
  V_i=A e^{-r_i^2/2\sigma^2}
  \label{eq:Vdip}
\end{equation}
where $\mathbf{r}_i$ is the position of site $i$, $A$ is the amplitude
and $\sigma$ the width of the perturbing potential.  We shall consider
negative $A$ values that create a dip in the confinement which in turn
induces a density bump at the lattice center of the form
$\delta n=\delta n_0 e^{-r_i^2/{{\tilde\sigma}^2}}$. The amplitude
$\delta n_0$ and the width $\tilde\sigma$ depend not only on $V_i$,
but also on the disorder configuration.  Once the density bump is
created, the additional Gaussian potential is turned off and the
system is let to evolve subject to the disordered potential. In a
circularly symmetric setup without any disorder, the initial density
bump would lead to a propagation of a ring-shaped fluctuation
characterized by its mean radius and transverse section. The evolution
of the mean radius measures the propagation speed and the size of the
transverse section, its broadening. However, since the different
momenta components scatter with the disorder at different speeds and
interfere with each other, the evolution of the density perturbation
become more complex, and the shape of the initial ring may be
lost. Therefore, the evolution will be mainly characterized by the
propagation speed and broadening of the angularly averaged density
fluctuation.

\begin{figure}
  \includegraphics[width=0.9\columnwidth,clip=true]{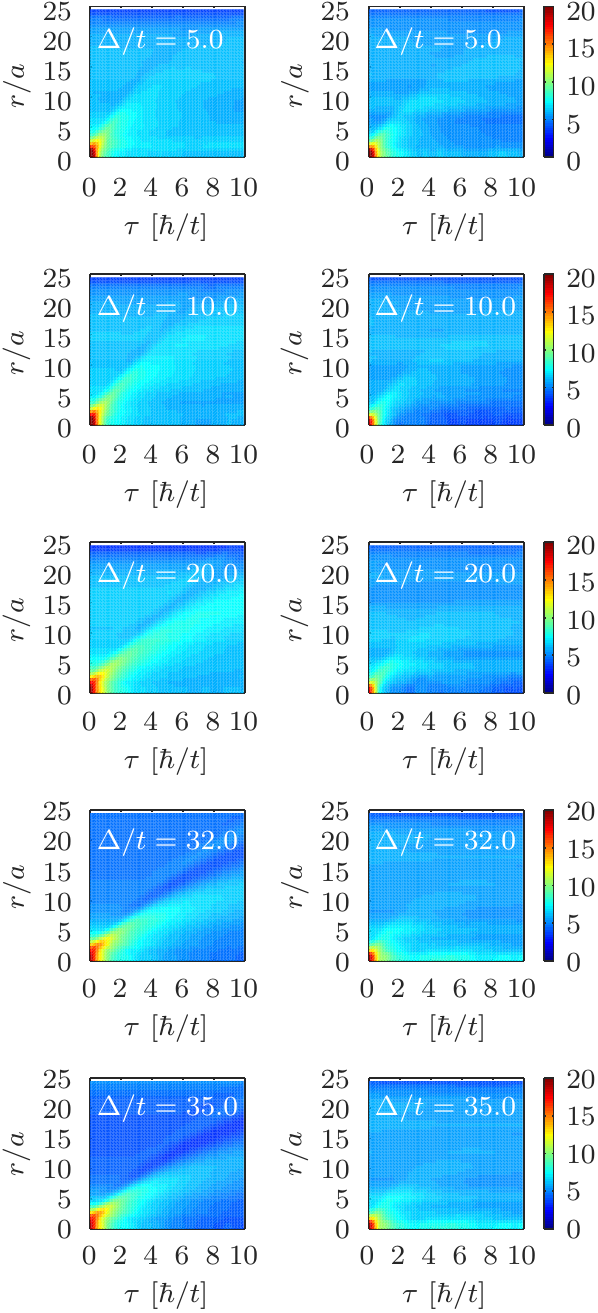}
  \caption{\label{fig3} Time evolution of the angularly averaged
    density $n(r,t)$ for an initial perturbation with $\sigma/a=1$ and
    $A/t=-5$ for different values of $\Delta$ for the case of a
    2D-DRDM lattice (left column) and of a UN-RAND system (right
    column). Each depicted density is obtained by averaging the
    density over 30 disorder realizations with $p=0.25$ for
    each model.}
\end{figure}

In Fig. \ref{fig3} we compare the time evolution of a density
perturbation in an UN-RAND lattice and in a 2D-DRDM one for $t'=3t$,
different values of $\Delta$ and $p=0.25$. We consider a system with a
weak interaction $U/t=10^{-2}$ and average number of particles per
site $\langle n_i\rangle = 5$ \cite{Capuzzi2015}. For any value of
$\Delta$ in the UN-RAND model, the density perturbation distorts as it
moves through the system; whereas in the 2D-DRDM, if $\Delta$ is close
to the resonant value $\Delta_{\text{res}}=32 t$, the density
perturbation propagates for a long time without a pronounced
dispersion.  This well-defined long-time density propagation indicates
that the density packet is not strongly deformed by the disorder
during its motion.

\begin{figure}
  \includegraphics[width=\columnwidth,clip=true]{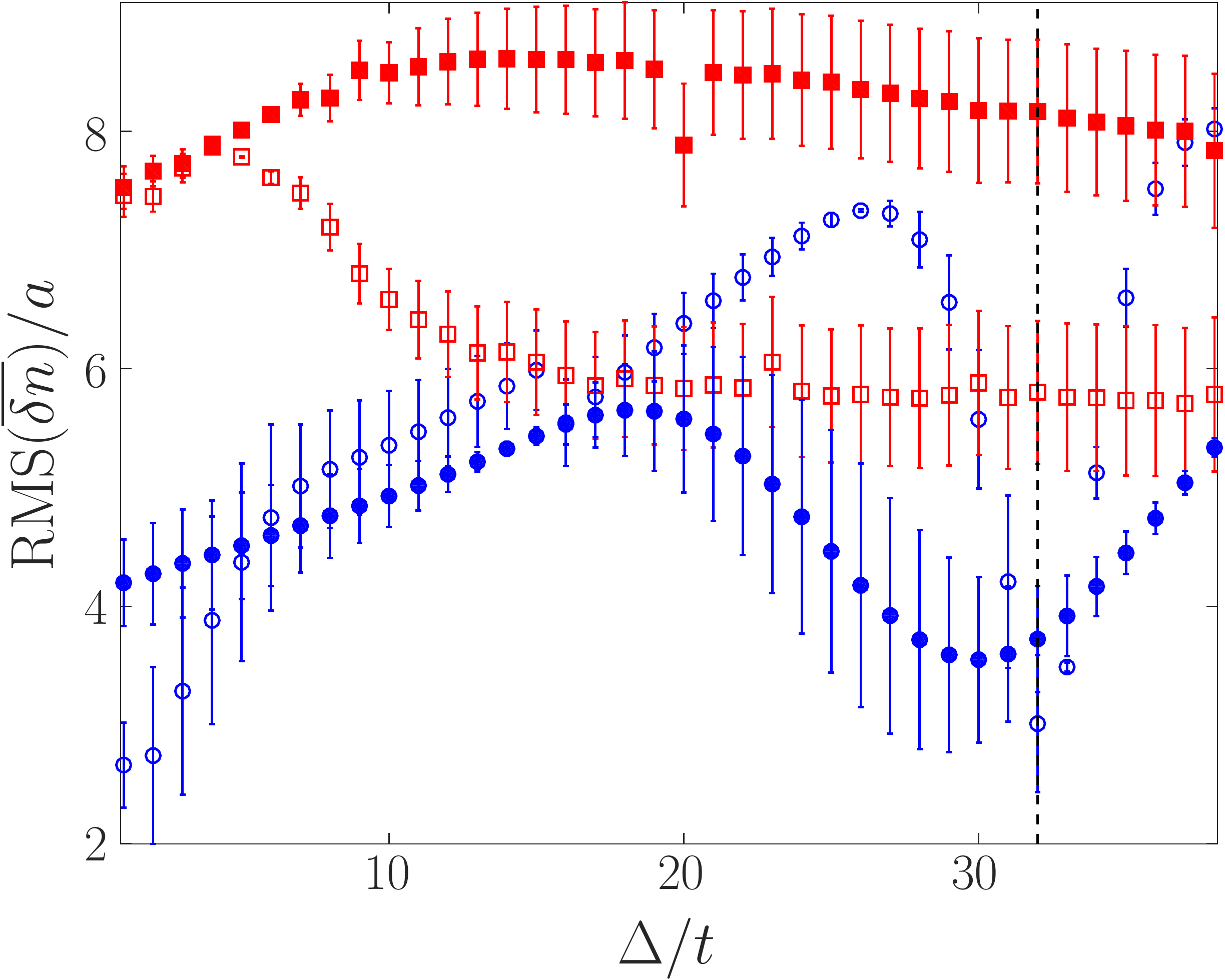}
  \caption{\label{fig4} RMS of average density perturbation for
    different values of $\Delta$ at time $\tau=9\hbar/t$, for the
    cases $U/t=10^ {-2}$ (empty symbols) and $U/t=10^ {-1}$ (full
    symbols).  Circles correspond to the 2D-DRDM model and squares to
    an UN-RAND one, both at $p=0.25$. The vertical dashed line
    indicates $\Delta_{\rm res}$.}
\end{figure}

\begin{figure}
  \includegraphics[width=\columnwidth,clip=true]{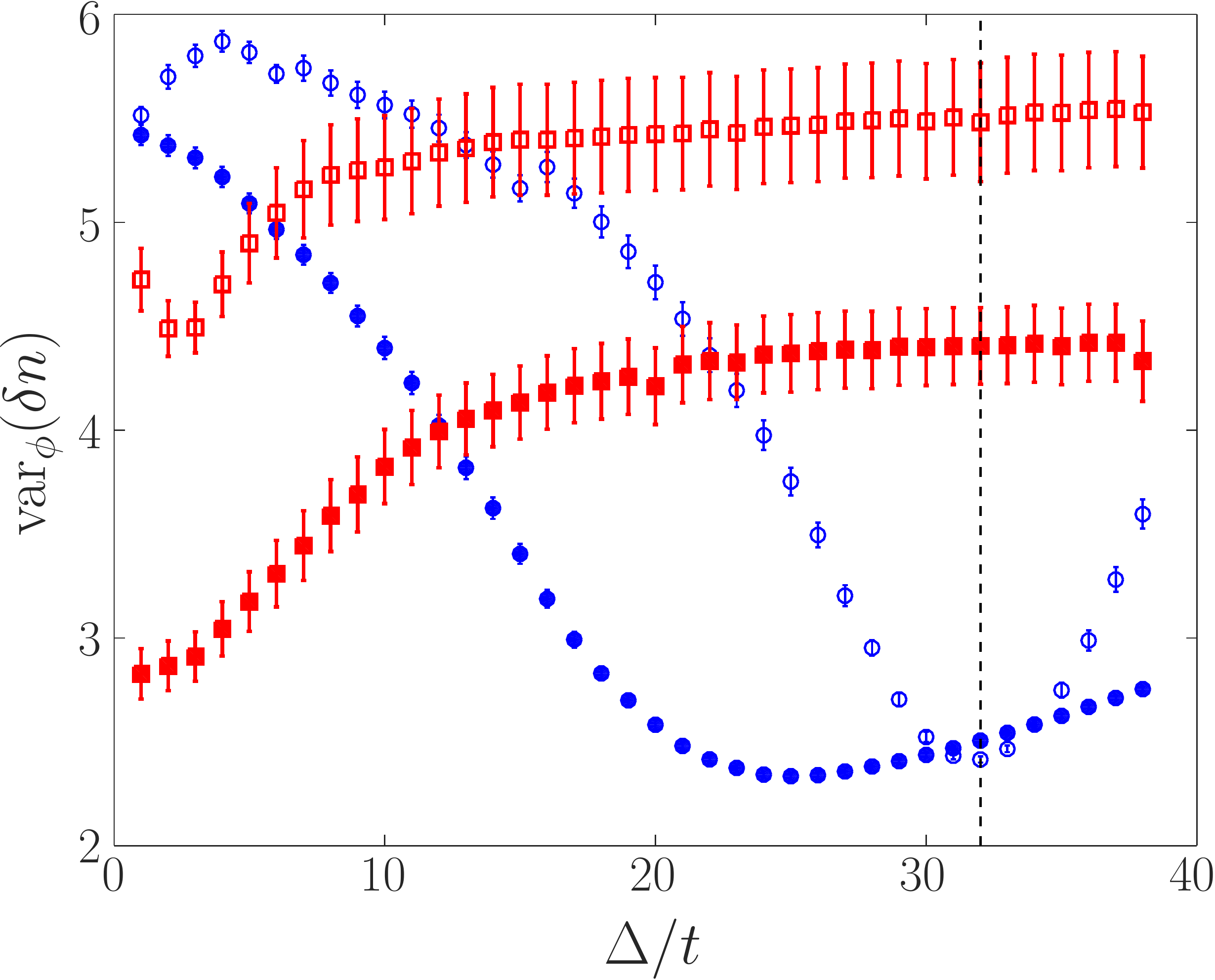}
  \caption{\label{fig5} Average angular variance of a density
    perturbation for different values of $\Delta$ at time
    $\tau=9\hbar/t$, for the cases $U/t=10^ {-2}$ (empty symbols) and
    $U/t=10^ {-1}$ (full symbols).  Circles correspond to the 2D-DRDM
    model and squares to an UN-RAND one, both at $p=0.25$. The vertical
    dashed line indicates $\Delta_{\rm res}$.}
\end{figure}

The deformation can be quantified by calculating the root-mean-square
(RMS) and the angular variance of the propagating density
fluctuations. The RMS is calculated as
$\mathrm{RMS}=\sqrt{\langle r^2\rangle-\langle r\rangle^2}$ where the
spatial averages are taken over the disorder-averaged fluctuations
$\overline{\delta n}$ as
$\langle F(r)\rangle=\int F(r)\overline{\delta n}^2/\int
\overline{\delta n}^2$.  On the other hand, the angular variance is
computed as
$\mathrm{var}_{\phi}(\delta n)=\sqrt{\int d\phi\, \delta n(\phi)^2
  -\left(\int d\phi\, \delta n(\phi)\right)^2}$ for the density
perturbation $\delta n(\phi)$ evaluated at the radius corresponding to
the location of the density peak. These magnitudes are measured at a
fixed final time of $\tau=9\hbar/t$ and shown in Figs. \ref{fig4} and
\ref{fig5}, respectively, as functions of $\Delta$ for two values of
the interaction strength: $U/t=10^{-1}$ (full symbols) and
$U/t=10^{-2}$ (empty symbols). The results in a 2D-DRDM and UN-RAND
lattice at $p=0.25$ are depicted in circles and squares, respectively.
For the case of the UN-RAND disorder, both these quantities are weakly
dependent on $\Delta$, while for the case of the 2D-DRDM disorder they
display a sharp minimum when the density perturbation remains
well-defined and does not spread out.  The position of the minimum
depends on the interaction strength: it corresponds to
$\Delta\simeq\Delta_{\text{res}}$ for the very weakly interacting gas
and moves to lower values by increasing the interactions
\cite{Capuzzi2015}.  It is worthwhile noticing that given the finite
size of the system representing experimental setups, it is not
possible to perform a detailed study of the long-time evolution of the
RMS, and thus to unambiguously distinguish between a wave-type and a
diffusive dispersion.

\subsection{Propagation in MP and BP lattices}
Aiming to understand the density perturbation propagation in the
disordered 2D-DRDM lattice, we study the dynamics in the MP and BP
lattices. The 2D-DRDM can be seeing as a MP lattice with dimerized
impurities or as a BP lattice with vacancies.  Analytical expressions
for the sound speed can be obtained if the gas is a pure Bose-Einstein
condensate (BEC).  The BEC limit corresponds to on-site coherent
states with Poissonian probability distribution in the Gutzwiller
ansatz
\begin{equation}
  |\Phi_{\text{BEC}}\rangle=\prod_i^{L\times L}\sum_{n_i}f_i(n_i)|n_i\rangle=
  \prod_i^{L\times L}\sum_{n_i}e^{-|\phi_i|^2/2}\dfrac{\phi_i^{n_i}}
       {\sqrt{n_i!}}|n_i\rangle,  
\end{equation}  
where $n_i=|\phi_i|^2$ is the on-site density, and $\phi_i$ is the
condensate wave function at site $i$.  In this case, the total energy
of the BEC reads
\begin{equation}
  E=-\sum_i(\mu-\varepsilon_i)|\phi_i|^2-
  \sum_{\langle ij \rangle}t_{ij}\phi_i\phi_j^*+
  \dfrac{U}{2}\sum_i|\phi_i|^4-\dfrac{UN}{2}.
\label{BECe}
\end{equation}
which gives rise to the equation of motion
\begin{multline}
  i\hbar\dfrac{\partial\phi_i}{\partial \tau}=\dfrac{\partial E}{\partial\phi_i^*}=
  -\sum_i (\mu-\varepsilon_i)\phi_i-\sum_{\langle ij\rangle}t_{ij}\phi_j
  \\+2U\sum_i|\phi_i|^2\phi_i.
\label{eq:BECdyn}
\end{multline}
The low-amplitude dynamics of the gas is governed by the collective
excitations of the system.  These are calculated through linearization
of the equation of motion (\ref{eq:BECdyn}) around a stationary
solution $\phi_i^0$ by setting
$\phi_i(t)=e^{-i\lambda\tau/\hbar}(\phi_i^0+u_ie^{i\omega\tau}+v_i^*e^{-i\omega\tau})$
where $u_i$ and $v_i$ are the amplitudes of the perturbation at site
$i$ and $\omega$ the corresponding frequency. The density fluctuation
can thus be written in terms of the collective mode $(u_i,v_i)$ as
$\delta n_i(t) = |\phi_i(t)|^2-|\phi_i^0|^2 \simeq
2\mathop{\mathrm{Re}}\left[e^{i\omega t}(\phi_i^0 u_i +
  v_i^*\phi_i^0)\right]$. The solution of the linearized dynamics
leads to the so-called Bogoliubov-deGennes eigenvalue equations
\cite{DeGennes1999,Leggett2001}.
\subsubsection{Bogoliubov spectrum in a MP lattice}
For the case of a MP lattice, where $t_{ij}=t$ for first neighbors, we
obtain the usual Bogoliubov-deGennes equations for the collective
modes in a lattice
\begin{equation}
  \begin{split}
    -\hbar\omega u_i=&-\mu u_i-t\sum_{\langle ij\rangle}u_j+2U\bar{n}u_i
    +U\bar{n}v_i\\
    \hbar\omega v_i=&-\mu v_i-t\sum_{\langle ij\rangle}v_j+2U\bar{n}v_i
    +U\bar{n}u_i\\
    \end{split}
 \end{equation} 
 where $\bar{n}=|\phi_i^0|^2=(\phi_i^0)^2=[(\phi_i^0)^*]^2$ is the
 density per site, and we have taken $\phi_i^0$ real.  Solving the
 above eigenvalue problem is straightforward and one obtain the energy
 spectrum of the collective excitations
 $\hbar\omega_k^{\rm MP}=\pm \sqrt{E_k^2+2E_k U\bar{n}}$, with
 $E_k=-2t\left[\cos(k_x a)+\cos(k_y a)\right]+4t$ the usual
 single-particle dispersion relation in a square lattice of spacing
 $a$. The low-$k$ slope of the collective spectrum defines the sound
 speed
 $c_s^{\rm MP}=\lim_{k\rightarrow 0}\partial\omega_k/(\partial
 k)=a\sqrt{2U\bar{n}t}/\hbar$
 \cite{Machholm2003,Liang2008a,Krutitsky2011}.
\subsubsection{Bogoliubov spectrum in a BP lattice}
For the case of a BP lattice, we can divide the system into two
sublattices of indices $i_a$, for sites hosting an impurity, and
$i_b$, for sites without impurities.  In this case, the
Bogoliubov-deGennes equations read 
\begin{equation}
  \begin{split}
    -\hbar\omega u_{i_a}=&(\Delta-\mu) u_{i_a}
    -t'\sum_{\langle ij\rangle}u_{j_b}
    +2U{n_a}u_{i_a}
    +U{n_a}v_{i_a}\\
    \hbar\omega v_{i_a}=&(\Delta-\mu) v_{i_a}-t'\sum_{\langle ij\rangle}v_{j_b}
    +2U{n_a}v_{i_a}
    +U{n_a}u_{i_a}\\
    -\hbar\omega u_{i_b}=&-\mu u_{i_b}
    -t'\sum_{\langle ij\rangle}u_{j_a}
    +2U{n_b}u_{i_b}
    +U{n_b}v_{i_b}\\
    \hbar\omega v_{i_b}=&-\mu v_{i_b}-t'\sum_{\langle ij\rangle}v_{j_a}
    +2U{n_b}v_{i_b}
    +U{n_b}u_{i_b}.\\
    \end{split}
\end{equation}
Here $n_a$ and $n_b$ are respectively the densities in each
sublattice, verifying the condition $n_a+n_b=2\bar{n}$.  The
lowest-energy band of the excitation spectrum is straightforwardly
calculated and reads
\begin{equation}
  \begin{split}
    &   \hbar\omega_k^{\rm BP}=\pm \left[\tilde{E}_k^2 +8(t')^2
      \dfrac{n_a^2+n_b^2}{n_an_b}
      +8t'U\sqrt{n_an_b}+2\times\right.\\
      &\left.\sqrt{\dfrac{(4t'\delta n)^2}{n_a^2n_b^2}
        +\tilde{E}_k^2\left(\dfrac{8\bar{n}^2}{n_an_b}\left(2(t')^2+
        \sqrt{n_an_b}t'U\right)+U^2\right)}\right]^{1/2},
\end{split}
  \end{equation}
  where $\tilde{E}_k=-2t'(\cos(k_x a)+\cos(k_y a))$ and
  $\delta n=n_b-n_a$.  We thus find that the sound speed for the
  bipartite lattice reads
\begin{equation}
  c_s^{\rm BP}=\frac{a}{\hbar}\sqrt{2t'(n_an_b)^{1/2}U\,\left[1+\frac{\delta n^2
  t'}{4 \bar{n}^2t'+(n_a n_b)^{3/2} U}\right]}.
\end{equation}
Due to the fixed average density $\bar{n}$, the sound speed in the
bipartite lattice varies with $\Delta$ through the variation of $n_a$
and $n_b$.  For vanishing $\Delta$ ($\delta n =0$), the expression of
$c_s^{\text{BP}}$ reduces to that of $c_s^{\text{MP}}$ with $t=t'$,
while for very large $\Delta$ it goes to zero as $n_a$ vanishes and
thus no perturbation can be transported.

\subsection{Density perturbation size effects}
\begin{figure}
  \includegraphics[width=0.95\columnwidth,clip=true]{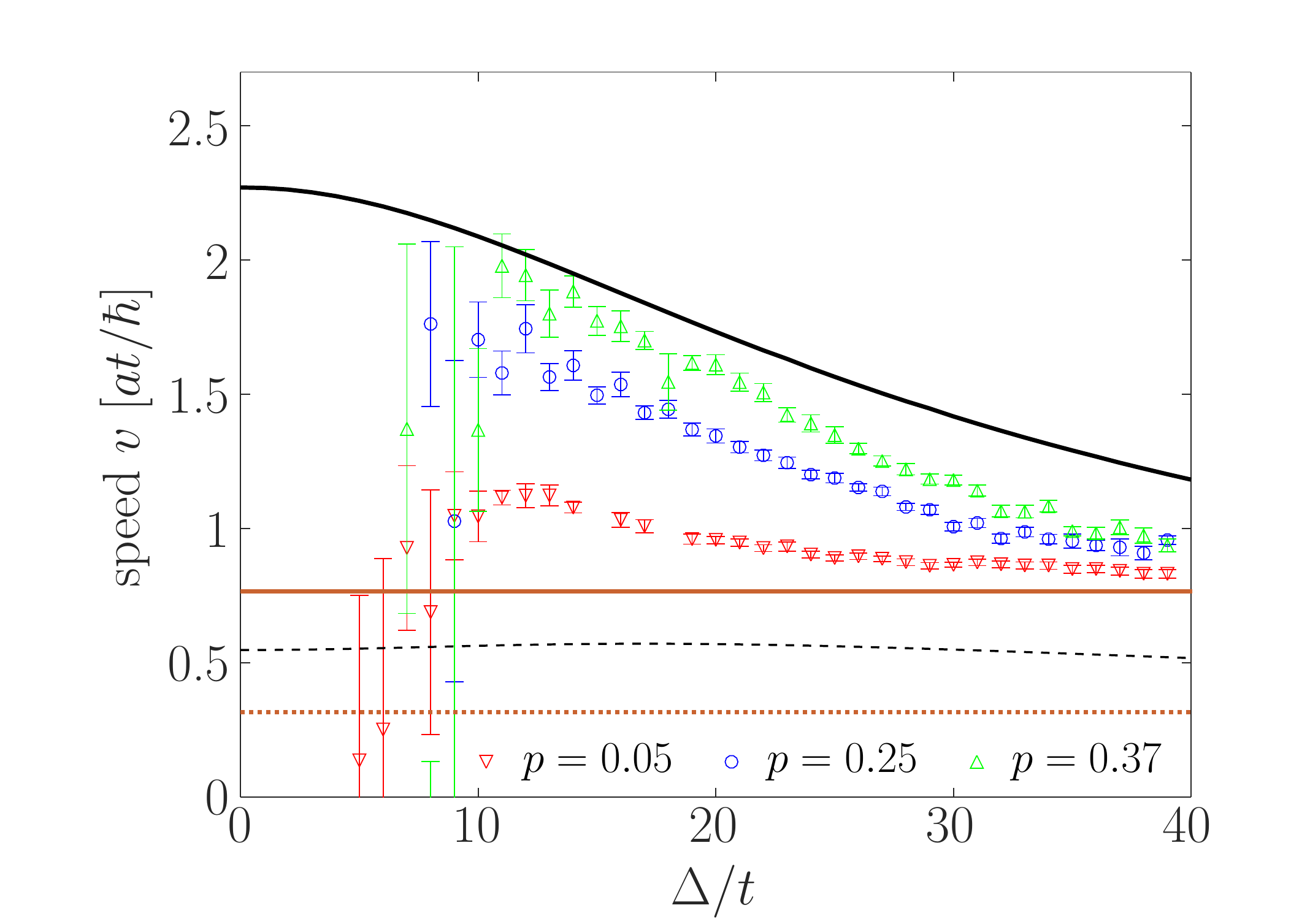}
  \includegraphics[width=0.95\columnwidth,clip=true]{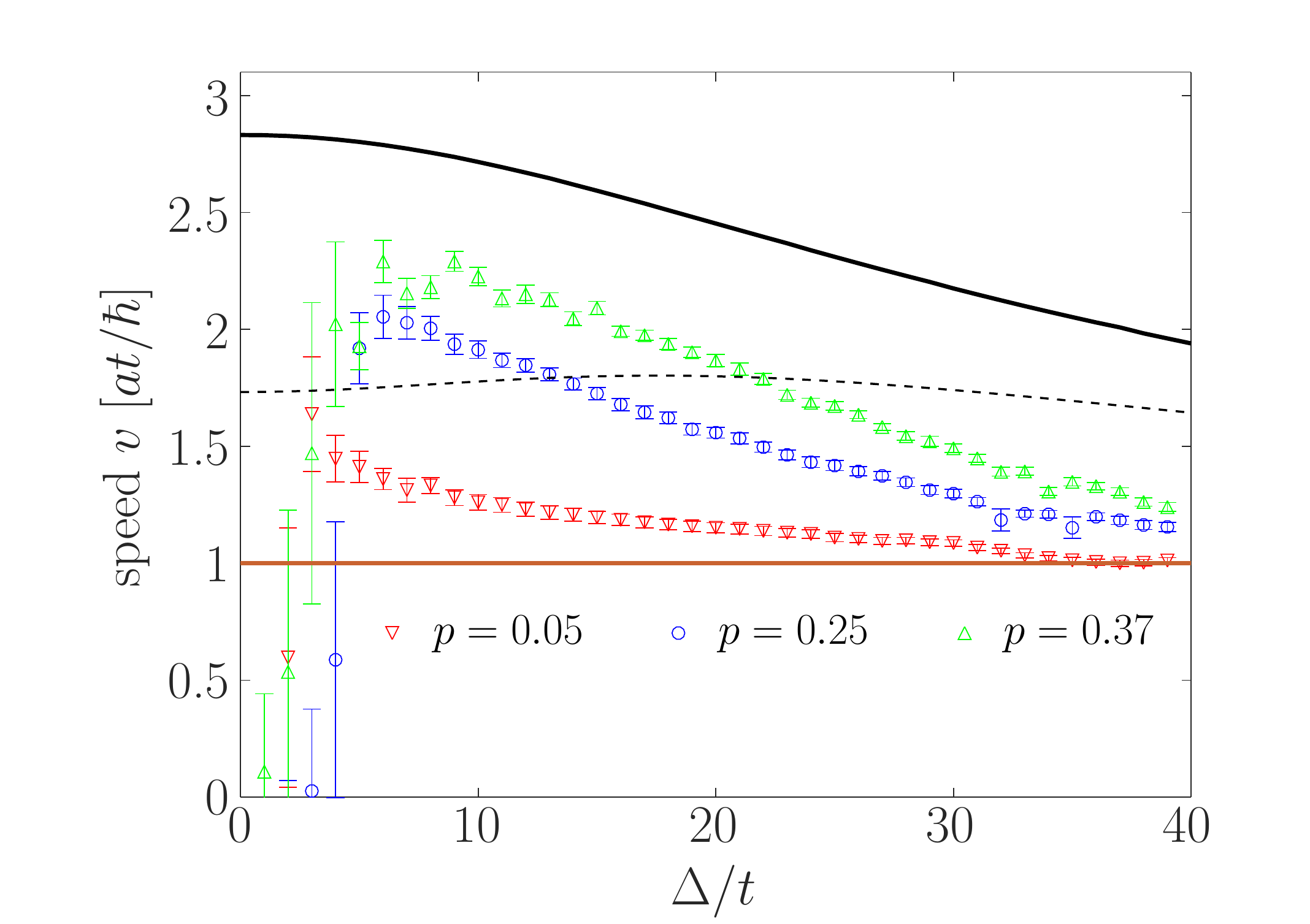}
  \caption{\label{fig6} Speed $v$ of the density perturbation in
    the 2D-DRDM lattice as a function of $\Delta$ for $U/t=10^{-2}$
    (top panel), and $U/t=10^{-1}$ (bottom panel). The different
    symbols correspond to different values of $p$. The solid black
    curves correspond to the BP propagation speed for
    $\tilde{\sigma}/a=2.2$ (see text), while solid brown curves
    correspond the MP propagation speed for $\tilde{\sigma}/a=2.4$
    (top panel) and $\tilde{\sigma}/a=4.0$ (bottom panel).  The dashed
    black (brown) curve corresponds to the sound speed for the BP
    (MP) lattice ($\tilde{\sigma}/a\rightarrow\infty$).  }
\end{figure}

When we excite the density wave as described above, the density
fluctuation propagates through the lattice with a speed that depends
on its spatial extent. For very large widths (small $k$) the
propagation speed $v$ coincides with the sound speed, while for tight
wave packets, larger-$k$ contributions have to be taken into account.
The group velocity at any $k$ contributes with a weight determined by
$\delta n(k)$, the density fluctuation in momentum space.  The actual
propagation speed can thus be written as
\begin{equation}
v\simeq c^{\rm MP,BP}=\int {\rm d}{\bf k} \,\dfrac{\partial\omega_k^{\rm MP,BP}}{\partial k}\,\delta n(k).
\label{eq:vsigma}
\end{equation}
Therefore, finite-$k$ corrections are more sizable for smaller $U$,
where the dispersion curves $\omega_k^{\rm MP,BP}$ bend more rapidly.
For the disordered lattices, the propagation speed has been extracted
from the position of the largest density peak during the evolution
when the dynamics is started by the Gaussian perturbation with
amplitude $A/t=-0.1$, and $\sigma/a=1$.  The choice of $\sigma/a=1$ is
determined by the experimental request that the density perturbation
has to be observable during the propagation along a finite-size
lattice.  The calculation of the finite-size velocities
$c^{\rm MP, BP}$ in Eq.\ (\ref{eq:vsigma}) have been performed for a
Gaussian wave packet of width $\tilde{\sigma}$. Due to the disorder,
$\tilde\sigma$ is usually larger than $\sigma$ and depends on the
system parameter $\Delta$ and the interaction strength $U$.  In
Fig. \ref{fig6}, we compare the propagation speed $v$ of a density
perturbation in a disordered 2D-DRDM lattice with $c^{\rm MP}$,
$c^{\rm BP}$, $c_s^{\rm MP}$ and $c_s^{\rm BP}$ for $U/t=10^{-2}$ (top
panel) and $U/t=10^ {-1}$ (bottom panel).  Given that for both
interaction strengths $\tilde\sigma/a\gtrsim 2$, in the weaker
interacting case $U/t=10^{-2}$, the size effects of the density
perturbation are quite important and the sound velocities
$c_s^{\rm MP}$ and $c_s^{\rm BP}$ largely underestimate $v$.  The
values of $\tilde{\sigma}$ used to calculate $c^{\text{MP}}$ and
$c^{\text{BP}}$ were fixed to those observed in the simulations in MP
($p=0$) and BP ($p=0.5$) lattices, respectively, for each value of
$U$. In the case of the BP lattice we have further chosen the
corresponding value of $\tilde{\sigma}$ at $\Delta/t=25$. For
$U/t=10^{-2}$, the density fluctuations do not propagate for
$\Delta/t\lesssim 10$, consistently with the behavior of the
noninteracting spectral function $\mathcal{A}$ shown in Fig.\
\ref{fig-spettro}.  For stronger interactions, the effect of the
finite size of the density wave packet diminishes as the linear range
of the dispersion curves $\omega_k$ extends to higher $k$. In this
case $c_s^{\rm MP}$ and $c_s^{\rm BP}$ correctly set the scale of the
data for $v$ (bottom panel of Fig. \ref{fig6}). As a result of the
interactions, the effect of the disorder is attenuated as demonstrated
by the reduction of the error bars for $\Delta/t\lesssim 10$,
indicating a well-defined propagation also at low $\Delta$. This can
be understood from the shift and broadening of the single-particle
energy resonance discussed in Sec.\ \ref{sec-model}.  The effects of
the resonance broadening on the properties of the ground state has
also been discussed in Ref.\ \cite{Capuzzi2015}.

\begin{figure}
  \includegraphics[width=0.9\columnwidth,clip=true]{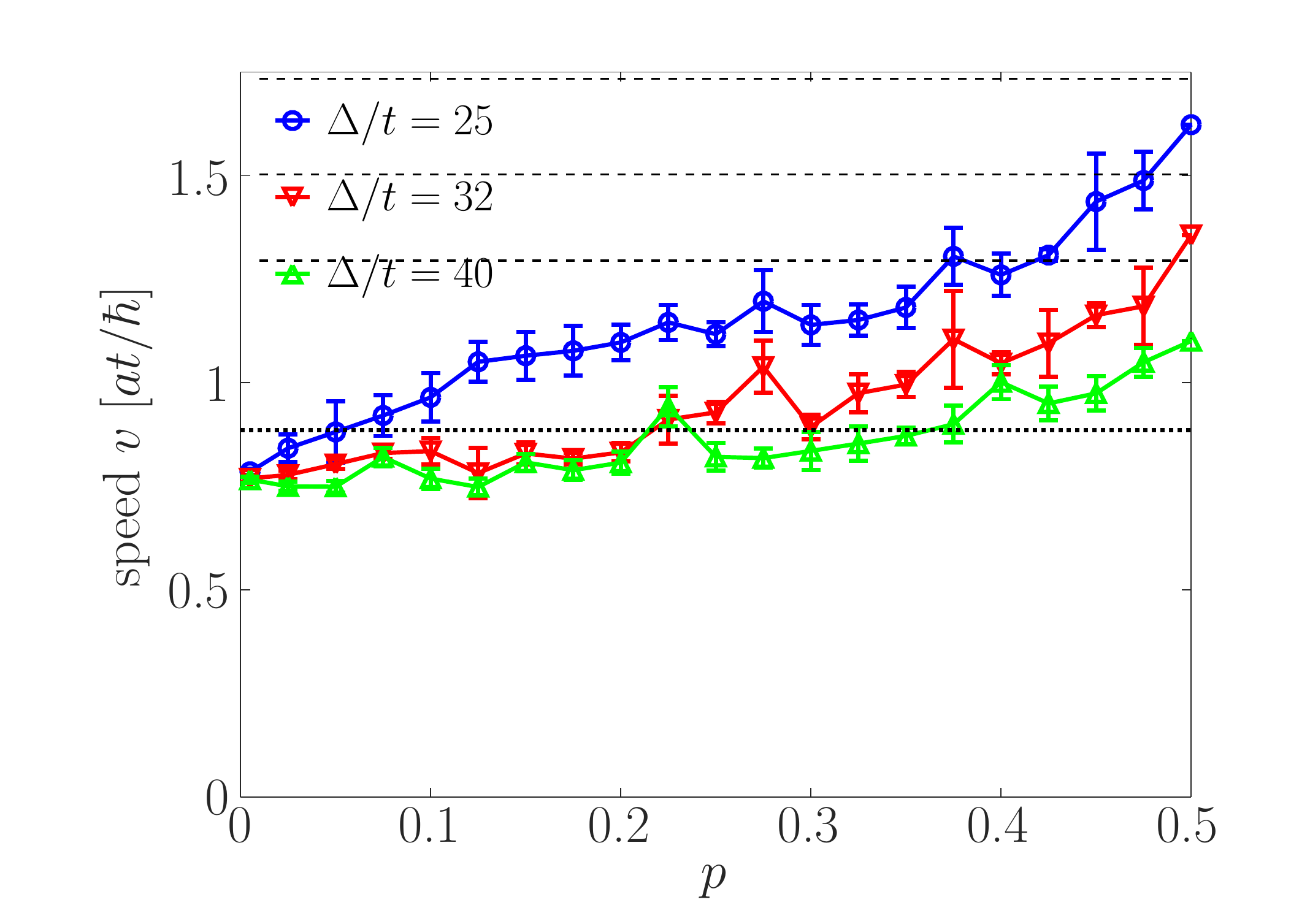}
  \caption{\label{fig7} Speed $v$ of the density perturbation in
    the 2D-DRDM lattice as a function of $p$ for $\Delta/t = 25, 32$ and
    $40$.  The black dotted-line corresponds to $c^{\mathrm{MP}}$ for
    $\tilde{\sigma}/a=2.4$, while the dashed-lines correspond to
    $c^{\mathrm{BP}}$ for the same values of $\Delta$ and
    $\tilde{\sigma}/a=2.2, 2.3$ and $2.5$ for $\Delta/t=25, 32$, and $40$,
    respectively.}
\end{figure}

The variation of the propagation speed with the percentage of
impurities $p$ is enlightened in more detail in Fig. \ref{fig7}, where
we plot $v$ as a function of $p$ for $\Delta/t= 25$, 32 and 40.
The data interpolate from $c^{\rm MP}$ at $p=0$ to $c^{\rm BP}$ at $p=0.5$.
%
From the density perturbation point of view, the 2D-DRDM lattice is
like an ordered lattice in between the MP and BP lattices.  The
dimerized impurities are not strictly-speaking ``transparent'' around
the resonance energy as it happens exactly at resonance.  Their
presence alters the propagation of density perturbations, but in the
same manner as if the impurities were orderly distributed.  The effect
of tuning the resonance energy to the ground-state energy is that, at
low-energies, the 2D-DRDM behaves like an ordered lattice
interpolating the MP and the PB lattices.


\section{Conclusions}
\label{sec:concl}
In this paper we have studied the propagation of an initial
ring-shaped density perturbation in a weakly interacting boson gas
confined on lattice and in the presence of localized disordered
impurities.  If the impurities are dimerized, and their resonance
energy corresponds to the ground-state energy, the density
perturbation propagates essentially without spreading out.  We found
that the speed of the density propagation is well defined even for a
tight density wave-packet (large wavevectors) and that its value
depends on the percentage of the impurities $p$, ranging from the MP
speed ($p=0$) to the BP one ($p=0.5$).  This means that the dimerized
impurities that are ``transparent'' at the ground-state energy, are
not strictly-speaking ``transparent'' at energies around the resonance
energy, and thus the density propagation depends on the impurities
presence.  The effect of the nearby resonance is that the system
behaves like that the dimerized impurities were orderly distributed,
for energies around the resonance energy.  If the resonance is far in
the spectrum, disorder correlations do not play any role, and the
density wave spreads out and does not propagate any more.
\begin{acknowledgments}
The authors acknowledge M. Gattobigio for useful discussions.  This
research has been carried out in the International Associated
Laboratory (LIA) LICOQ.  P. C. acknowledges partial support from
CONICET and Universidad de Buenos Aires through grants PIP
11220150100442CO and UBACyT 20020150100157, respectively.
\end{acknowledgments}

\end{document}